\documentclass[draftcls,11pt, onecolumn]{article}

\usepackage[left=3cm,right=3cm, bottom=3cm, top=3cm]{geometry}
\usepackage[utf8x]{inputenc}
\usepackage[american]{babel}
\usepackage{pstool}
\usepackage{amsmath,amsfonts,amssymb}
\usepackage{mathabx,fixmath}
\usepackage{algorithm}
\usepackage{algorithmic}
\usepackage{psfrag}
\usepackage{epstopdf}

\usepackage{cite,color}
\renewcommand{\a}{\mathbold{a}}

\renewcommand{\r}{\mathbold{r}}
\renewcommand{\v}{\mathbold{v}}
\newcommand{\w}{\mathbold{w}}
\newcommand{\x}{\mathbold{x}}

\newcommand{\A}{\mathbold{A}}

\newcommand{\T}{\mathbold{T}}

\newcommand{\W}{\mathbold{W}}

\newcommand{\bnu}{\mathbold{\nu}}

\newcommand{\btheta}{\mathbold{\theta}}

\newcommand{\Reals}{\mathbf{R}}

\newcommand{\transp}{\mathsf{T}}
\newcommand{\trace}{\mathrm{tr}}

\renewcommand{\thefootnote}{\arabic{footnote}}
\newtheorem{remark}{Remark}

\begin{document}
%
\title{Consistent Sensor, Relay, and Link Selection\\ in Wireless Sensor Networks\let\thefootnote\relax\footnote{The work in this paper was supported by Spanish Government grant TEC2014-52289.}}

	

\author{
Roc\'{\i}o~Arroyo-Valles\footnote{
R.~Arroyo-Valles was with the Dept. of Teor\'ia de la Se\~{n}al y Comunicaciones, Universidad Carlos III de Madrid, Avda. de la Universidad 30, Legan\'es, 28911,Madrid, Spain. She is now with the European Patent Office,
2280 HV Rijswijk, The Netherlands. E-mail: marrval@tsc.uc3m.es.}, 
Andrea~Simonetto\footnote{
Andrea Simonetto was with the applied mathematics department, Universit\'e catholique de Louvain, Louvain-la-Neuve, Belgium. He is now with the optimization and control group of IBM Research Ireland. E-mail: andrea.simonetto@ibm.com}, 
and Geert~Leus\footnote{
Geert Leus is with the Faculty of Electrical Engineering, Mathematics and Computer Science, Delft University of Technology, 2628CD Delft, The Netherlands. E-mail: g.j.t.leus@tudelft.nl
}}



\maketitle

\begin{abstract}

In wireless sensor networks, where energy is scarce, it is inefficient to have all nodes active because they consume a non-negligible amount of battery. In this paper we consider the problem of jointly selecting sensors, relays and links in a wireless sensor network where the active sensors need to communicate their measurements to one or multiple access points. Information messages are routed stochastically in order to capture the inherent reliability of the broadcast links via multiple hops, where the nodes may be acting as sensors or as relays. We aim at finding optimal sparse solutions where both, the consistency between the selected subset of sensors, relays and links, and the graph connectivity in the selected subnetwork are guaranteed. Furthermore, active nodes should ensure a network performance in a parameter estimation scenario. Two problems are studied: sensor and link selection; and sensor, relay and link selection. To solve such problems, we present tractable optimization formulations and propose two algorithms that satisfy the previous network requirements. We also explore an extension scenario: only link selection. Simulation results show the performance of the algorithms and illustrate how they provide a sparse solution, which not only saves energy but also guarantees the network requirements.

\end{abstract}

%
%


\section{Introduction}
\label{Introduction}


Nowadays, wireless sensor networks are developed to provide fast, cheap, reliable, and scalable hardware solutions for a large number of industrial applications, ranging from surveillance \cite{Biswas2006, Raty2010} and tracking \cite{Songhwai2007, Liu2007a} to exploration \cite{Sun2005, Leonard2007}, monitoring \cite{Corke2010, Sun2011}, and other sensing tasks \cite{Arampatzis2005}. From the software perspective, an increasing effort is spent on designing algorithms that can provide high reliability with limited computation, communication, and energy requirements for the sensor nodes.
 
In this paper, we consider a network of battery-powered sensors that take measurements related to some important environmental parameter and that need to communicate their measurements to one or multiple access points (APs), or sinks, which are responsible for processing the gathered information. Communication with the APs is achieved through multihop routes defined via a connectivity graph which considers the sensors' communication range.
 
Resources (mainly energy) in this network are scarce so it is inefficient to have all sensors active. Some sensors may not be informative enough and hardly contribute to achieve a minimum desired network performance; nonetheless, if active, they would consume a non-negligible amount of resources. Moreover, communication efforts are among the most energy demanding tasks in wireless sensor networks~\cite{Raghunathan02} and they should be minimized by properly selecting not only the suitable sensors but also the proper active links. Knowledge of the network topology should be exploited in order to make a better selection of the links that are in charge of conveying the information because information may be degraded over long distances and transmissions should be avoided to reduce energy expenditure.

With the reduction of energy expenditure in mind, in this paper we consider a distributed estimation scenario in wireless sensor networks, where each sensor takes local measurements of a phenomenon of interest at a particular rate and communicates them in a multihop way to one or multiple APs. In this scenario, we study the problem of judiciously and consistently selecting the optimal minimum set of sensors \emph{and} links that ought to be active in the network, so that a prescribed network performance (e.g., the mean squared error of the estimation of the parameter of interest) \emph{as well as} graph connectivity among the selected active sensors are guaranteed. Only the measurements taken by the active sensors must be reported back to the APs via the active sensors and active links. This is the reason for requiring graph connectivity among the selected active sensors. Moreover, the optimal sensing rates supported by the active sensors are calculated. We analyze the problem from a stochastic point of view, where information messages are routed stochastically thereby capturing the inherent reliability of the broadcast wireless links.

\subsection{State of the art}

The concept of sensor selection has been extensively studied in the context of parameter and state estimation. The resulting minimum cardinality combinatorial problem has been tackled by using different tools, from convex relaxations, e.g., \cite{Joshi09, Chepuri15, Jamali-Rad15}, to sub-modularity~\cite{Shamaiah2010, Liao09, Naeem09} and frame theory~\cite{Ranieri14,Zhao12}. These tools have their pros and cons. 
More along the lines of this paper, in \cite{Liu15} not only is the best subset of sensors selected that communicate with the fusion center but also the collaboration scheme that allows each sensor to combine its raw measurements with those coming from other sensors according to certain weights. 




Stochastic routing in multihop networks has been introduced in the literature in order to cope with the random nature of wireless links \cite{Sivrikaya09, Ribeiro08}. Transmissions are based on a reliability matrix, where each element of the matrix reflects the probability of satisfactorily transmitting and receiving a message between two given sensors.
In \cite{Zavlanos13}, the authors define the concept of connectivity within a context of mobile robotic networks in terms of communication rates, and based on this definition, the authors propose a distributed algorithm to find the optimal operating points of wireless networks when the link metric is the link reliability.
The work of~\cite{Shah2014} considers the problem of optimizing the routing and sensor selection given a total budget constraint. Yet, the approach presented in~\cite{Shah2014} is heuristic and divides the estimation and routing problems, by tackling them in two separated phases, which could cause additional suboptimality of the solution. 


Often times, a distinction is made between sensor and relay nodes.  Relay nodes help the source nodes (sensing nodes) in forwarding the messages to the APs: they receive a message from the source nodes, process it and forward it towards the intended APs. Relaying is especially beneficial when there is no line-of-sight path between the source and the destination. This distinction between sensor types may be motivated, for instance, by economical reasons (relay devices may be cheaper than sensors given that their functionality is more limited), or by design prerequisites (sensors need to achieve a certain performance while relays do not because they are only limited to forwarding the information). Previous state-of-the-art works are only based on proposing relay selection schemes (e.g., \cite{Ibrahim08, Lo09}, and references therein): given a source sensor and a sink, they try to choose the best relays among a collection of available ones based on different criteria. Other works are aimed at optimally placing wireless relay nodes and sinks \cite{Bhattacharya14}.

\subsection{Our contributions}

All the aforementioned state-of-the-art works either face the sensor selection problem \emph{or} the stochastic routing, but what has never been addressed in the literature before is the challenge of jointly selecting the optimal minimum set of active sensors (and their corresponding sensing rates) which satisfies a prescribed estimation performance metric and consistently finding the optimal multihop routes so that the selected subgraph is connected. Hence, in this paper we do not focus on devising new methods to solve selection problems or on comparing them, instead we are mainly interested in formulating a stochastic framework for consistent sensor and link selection. Even the closest prior work \cite{Liu15}, which is a ``dual'' problem w.r.t. ours, differs from this paper in several ways: in \cite{Liu15} all sensors can directly communicate with the fusion center (i.e., it is not a multihop scenario so the graph connectivity is not a problem), communication links are established based on inter-sensor collaboration before transmitting a processed message to the fusion center, and the optimal transmission rates of transmitting sensors are not determined.

The problem at hand becomes even more challenging when there is a distinction between sensor and relay nodes. In a scenario where there are the two types of nodes, we want to consistently determine which of the nodes, placed at well-determined positions, should play the role of sensors (and hence their sensing rate should be determined) and which ones the role of relays while guaranteeing both a prescribed network performance and connectivity in the selected subgraph. To find an optimal solution, a joint source and relay selection should be performed, which implicitly implies to activate suitable links.

The main contributions of this paper can be summarized as follows. 

1) From a stochastic point of view and in a multihop scenario, we formulate a tractable optimization problem to select consistently the optimal subsets of sensors (with their sensing rates) and links that guarantee both, a required network performance and graph connectivity in the selected subnetwork. To solve this problem, we propose a sparsity-aware algorithm based on a convex relaxation (Section \ref{ProblemFormulation} to Section \ref{SimulationResults1}).
 
2) The previous framework is also well-suited for the joint selection of sensors, relays and links (which is not the case for other approaches in the literature, e.g.,~\cite{Shah2014}). Under a slight modification of the previous optimization problem and applying a convex relaxation technique, we propose another sparsity-aware consistent sensor-relay-and-link selection algorithm. This algorithm assigns the optimal sensing rates to the active sensors and ensures network connectivity as well as a prescribed network performance (Sections \ref{Relays} and \ref{SimulationResults2}).

3) Finally, we also extend the work to a special case where only link selection is considered (Section \ref{SpecialCasesExtensions}).

Contributions 1)-3) rely on a reformulation of the problems as $\ell_1$ convex optimization problems. This allows for efficient and well performing algorithms. Different approaches, e.g.~\cite{Dai2011}, would yield more complex problem formulations, which rely on dedicated non-convex solvers. This is avoided here. In addition, based on the fact that a $\ell_1$ relaxation is leveraged, distributed algorithms can be envisioned (See Remark~\ref{rem.distributed}).  

\begin{figure}
\includegraphics[width=1\textwidth]{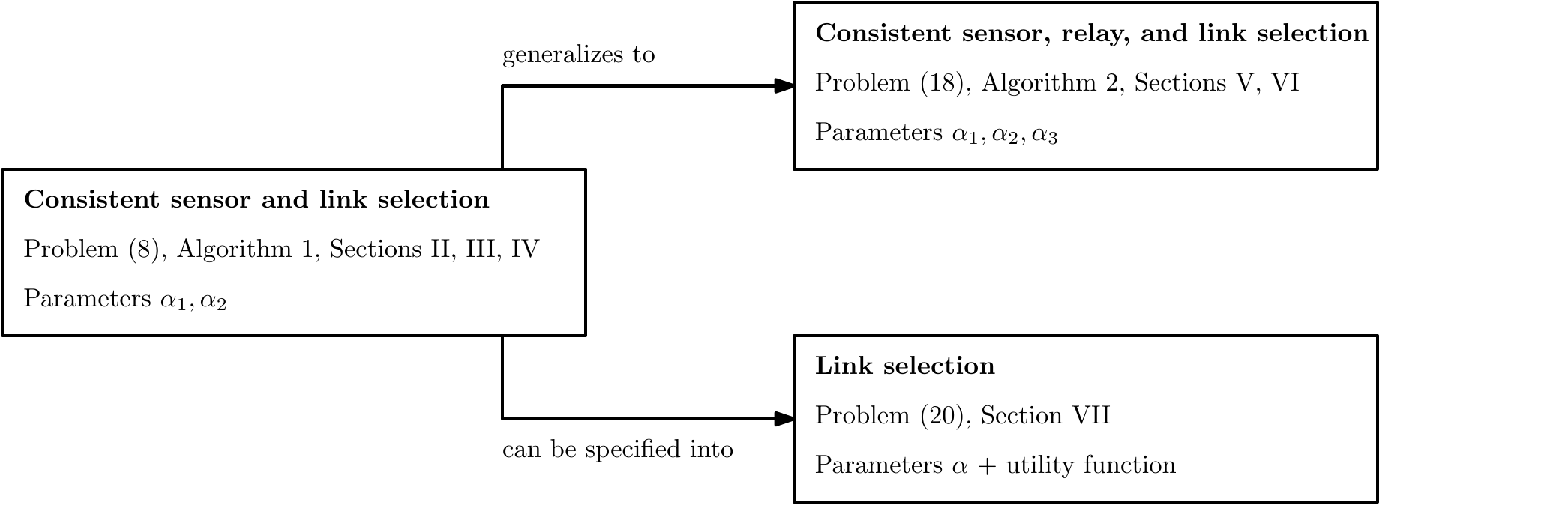}
\caption{General structure of the paper with the three problem formulations and relations among them.}
\label{fig.schema}
\end{figure}


Numerical simulation results support our claims and illustrate a satisfactory performance of the proposed algorithms. As a last note, we highlight that the presented algorithms are exposed in a static framework, i.e., given a certain network, we provide a selection strategy. Yet, they could be implemented in a dynamical way, by repeating their execution, so to balance the energy level of the active and non-active sensors and relays (See Remark~\ref{rem.energy}).


\vskip2mm

\textbf{Notation.} Notation is where possible standard: we indicate with boldfaced small letters, such as $\x$, real vectors, whereas capital boldfaced letters, e.g. $\A$, represent real matrices. Vector $p$-norms are indicated with $\|\cdot\|_p$, while $p$-norms for matrices are intended \emph{element-wise}, e.g., $\|\A\|_1$ is the sum of the absolute values of the elements of the matrix $\A$. Pseudo-norms, such as the $0$-norm, follow the same notation.  

\section{Problem Formulation}
\label{ProblemFormulation}

\textbf{High-level problem description. }
In this paper, we are facing the problem of consistently selecting the smallest subset of sensors and links out of all available ones such that a certain performance measure and network connectivity (which ensures a path from the active sensors to the APs) is guaranteed. The motivation behind selecting a low number of sensors (and subsequently, an appropriate reduced amount of links) comes from the need of minimizing the economical and communication costs in wireless sensor networks. Clearly, this saving should not jeopardize the performance or the network connectivity. Communication between the active sensors and the APs as well as a network performance must be guaranteed.

\vskip2mm

We consider a static wireless sensor network composed of $J$ sensor nodes and $K$ access points (APs) or sinks. At this point, we don not consider any relays yet. 
We denote with $\mathcal{V}=\{1,2,..., J, J+1, ..., J+K \}$ the set of sensors and access points, where $i\in\mathcal{V}_{\textrm{s}} = \{1,...,J\}$ are the indexes corresponding to the sensor nodes  and $i\in\mathcal{V}_{\textrm{AP}}=\{J+1, ..., J+K\}$ are the indexes corresponding to the APs. The network topology is determined by the physical locations of the sensors and APs, collected in the stacked vector $\x=[\x_1^\transp,...,\x_J^\transp, \x_{J+1}^\transp,\dots,\x_{J+K}^\transp ]^\transp$, where the vector $\x_i$ indicates the position of sensor or AP $i$. 

\subsection{Communication Network}

Sensors need to communicate their measurement to the APs in a multi-hop fashion (due to energy/power constraints). An important feature of this paper is that we can only use active sensors to transmit messages. We model the communication quality among sensors and APs using a link reliability metric, denoted as $R_{ip} := R(\|\x_i-\x_p\|)$, which represents the probability that sensor $p$ (if $p \leq J$) or an AP (otherwise) receives successfully a message sent from sensor $i$. We model this probability as a smooth non-increasing function with compact support, and in particular, $R(0) = 1$ and $R(d) = 0$ for all $d\geq \bar{d}$, for a predefined cut-off distance $\bar{d}$. 

The link reliability metric induces a specific undirected communication graph on the wireless sensor network: whenever $R_{ip}$ is nonzero, there is a possible link between sensor $i$ and sensor or AP $p$. We describe this communication graph in terms of the edge set $\mathcal{E}$, given by $\mathcal{E} = \{ (i,p), i \in \mathcal{V}_{\textrm{s}}, p \in \mathcal{V} | i\neq p, R_{ip} > 0\}$, and we denote the graph as $\mathcal{G} = (\mathcal{V}, \mathcal{E})$.

\subsection{Sensing}

Sensors take measurements of a parameter $\btheta \in \Reals^m$, $m \ll J$, according to the linear measurement model, 
\begin{equation}
y_i = \a_i^\transp \btheta + n_i,\quad i\in\mathcal{V}_{\textrm{s}},  
\end{equation}
where the vectors $\a_i\in\Reals^m$ represent the regressors, while $n_i$ is a Gaussian noise term with mean 0 and covariance $\sigma_i^2$. 
Sensor $i$ acquires measurements $y_i$ with a rate $r_i \bar{r}_{i}$ (we assume that the maximum relative rate $\bar{r}_{i}$ is known and fixed, while the relative rate $r_i \in [0,1]$ is a design parameter). If $r_i=0$, the node will not take any measurements and will not be active.

As we mentioned, the collected measurements need to be communicated back to the APs in a multi-hop fashion. The APs are in charge of combining the measurements $y_i$, coming from different sensors at different rates, to estimate the value of the parameter $\btheta$. The \emph{quality} of the estimate can be evaluated a priori based on which sensors are measuring (more specifically their regressors $\a_i$ and noise variances $\sigma_i^2$) and their rates. Examples of such quality metrics are rate versions of the mean square error (MSE), the worst case error variance, or the volume of the confidence ellipsoid~\cite{Joshi09}. For instance, if we select the MSE-rate as quality metric and assume that the noise experienced at different sensors is uncorrelated, then we would have 
\begin{equation}
f(\r) := \trace\Big(\sum_{i\in\mathcal{V}_{\textrm{s}}} r_i \bar{r}_{i} \a_i \a_i^\transp /\sigma_i^2\Big)^{-1}\!\!\!,
\label{CostFunction}
\end{equation}
where we have collected the relative rates in $\r = [r_1, \dots, r_J]^\transp$. Remark that if a sensor is not active, its relative rate is zero. The higher the value of $f(\r)$, the higher the MSE-rate of the estimate, and vice versa. {Other types of function $f(\r)$ can be found in~\cite{Joshi09, Jamali-Rad15}, both for uncorrelated and correlated noise}. In order to keep the presentation as general as possible, we will not specify which quality metric we select: we will simply write the metric as the function $f(\r)$.     

\subsection{Connectivity Modeling}

Before formalizing the problem mathematically, we need to introduce how we will model the communication links and the induced connectivity. In this paper, we use a stochastic point of view and we use the stochastic routing framework of \cite{Zavlanos13}. 

In our multihop wireless network, messages will be routed stochastically, i.e., sensor nodes select a neighbor, either a sensor or an AP, to forward the message according to a certain probability. A set of variables $T_{ip} \in [0,1]$ will denote the probability that node $i$ selects node $p$, either a sensor or an AP, as a destination of the transmitted messages. In this sense, the variables $T_{ip}$ can be seen as the probability that node $i$ selects the link that joins sensors $i$ and $p$. The matrix $\T$, of size $J \times (J+K)$, gathers all these probability values. Further, the matrix $\T$ needs to satisfy a certain number of constraints. First, if either one of the sensors $i$ or $p$ is not active, then $T_{ip}$ must be zero: this models the fact that if a sensor is not active then it cannot send or receive messages. This can be formulated as 
\begin{equation}\label{eq.Tminw}
T_{ip} = 0 \quad \textrm{iff}\quad r_i r_p = 0, \quad i\in\mathcal{V}_{\textrm{s}}, p \in \mathcal{V},
\end{equation} 
since $T_{ip}$ will be nonzero if and only if both $r_i$ and $r_p$ are nonzero, meaning that the sensors are active (we can fix $r_p = 1$ for APs, without loss of generality). Second, since we are dealing with link probability values, the sum of all link probabilities of an active sensor should be at most $1$:
\begin{equation}\label{eq.prob1}
\sum_{p\in\mathcal{V}} T_{ip} \leq 1, \quad i\in\mathcal{V}_{\textrm{s}}.
\end{equation} 
We notice that we can use $i\in\mathcal{V}_{\textrm{s}}$ in the condition~\eqref{eq.prob1}, since non-active sensors have $T_{ip} = 0$ due to condition~\eqref{eq.Tminw}, and therefore \eqref{eq.prob1} is automatically satisfied.

To complete the formulation, we want to ensure the delivery of messages to the APs, which is achieved by guaranteeing network connectivity among the active sensors and APs. To that aim, let $R_0$ be the transmission rate of the sensors. Then the effective transmission rate in the active link between nodes $i$ and $p$ is $R_0 R_{ip}$ (recall that $R_{ip} := R(\x_i,\x_p)$ is the link reliability between sensors or APs). We consider normalized rates by making $R_0=1$, and we further assume that all sensors have the same transmission rate $R_0$, which is an easy-to-lift constraint. 

Each sensor stores messages in a queue between the generation or arrival from other sensors and their transmission. An active sensor $i$, apart from generating messages locally at rate $r_i \bar{r}_i$, also receives messages from other sensors $p$ with an active link $T_{pi}R_{pi}$. Thus, the aggregate rate at which messages arrive at sensor node $i$ is 
\begin{equation}
\label{r_in}
r_i^{\textrm{in}}= r_i \bar{r}_i + \sum_{p\in\mathcal{V}_{\mathrm{s}}} T_{pi}R_{pi}.
\end{equation}
In a similar way, the rate at which sensor $i$ sends messages to other nodes $p$, which may be sensors or APs, is given by
\begin{equation}
\label{r_out}
r_i^{\textrm{out}} = \sum_{p\in\mathcal{V}}T_{ip}R_{ip}.
\end{equation}
%
If we consider that the average rate at which messages leave the sensor's queue is higher than the rate at which messages arrive at a sensor, i.e., $r_i^{\textrm{out}} \geq r_i^{\textrm{in}}$, i.e., 
\begin{equation}\label{eq.flow}
r_i \bar{r}_i + \sum_{p\in\mathcal{V}_{\mathrm{s}}} T_{pi}R_{pi} \leq  \sum_{p\in\mathcal{V}}T_{ip}R_{ip}, \quad i \in \mathcal{V}_{\textrm{s}},
\end{equation}
then the queue empties often with probability one and there is an almost sure guarantee that the messages are delivered to the AP \cite{Zavlanos13} ({a formal statement of this fact will be given in the following}). 

\textbf{Problem statement}: Given the measurement model for the different sensors and a prescribed performance measure value $\gamma>0$, we want to find the relative rates $\r \in [0,1]^J$, which select the minimum subset of sensors, and the probabilistic routing matrix $\T \in [0,1]^{J \times (J+K)} $, which selects the minimum subset of links, so that the performance measure $f(\r) \leq \gamma$ is satisfied and the messages are delivered to the APs. This can be stated as
\begin{subequations}
\label{Problem.nonconvex} 
\begin{align}
 & \underset{\r,\T}{\text{minimize}}  
 & & \alpha_1 \|\r\|_0 + \alpha_2 \| \T \|_0 \\
 & \text{subject to} & & r_i \in [0,1],\, T_{ip} \in [0,1], \quad i\in\mathcal{V}_{\textrm{s}}, \, p \in\mathcal{V}  \\
& 
&& \eqref{eq.Tminw}, \eqref{eq.prob1}, \eqref{eq.flow} \\ 
& 
&& f(\r)  \leq \gamma, 
\end{align}
\end{subequations}
where the non-negative scalars $\alpha_1$ and $\alpha_2$ determine the importance of the sensors and the links. If $\alpha_1 = 0$, then the problem becomes link selection with stochastic routing, while for $\alpha_2 = 0$, the problem is sensor selection. We denote as $(\r^*, \T^*)$ any optimal couple determined by the solution of the problem~\eqref{Problem.nonconvex}. 

We can readily notice that \eqref{Problem.nonconvex} is a nonconvex program, which makes finding any optimal couple $(\r^*, \T^*)$ computationally expensive in practice. In this paper, we are interested in finding an approximate solution of~\eqref{Problem.nonconvex} by a suitable convex relaxation.

\section{Convex relaxation}
\label{Convexrelaxation}

We relax the nonconvex program~\eqref{Problem.nonconvex} by substituting the $\ell_0$-pseudo norm, with the $\ell_1$ norm, and by substituting the nonconvex constraint~\eqref{eq.Tminw} with the convex surrogate
\begin{equation}\label{eq.Tminw_cvx}
T_{ip} \leq \min\{r_i, r_p\}, \quad i\in\mathcal{V}_{\textrm{s}}, p \in \mathcal{V}.
\end{equation} 
These operations transform the original problem~\eqref{Problem.nonconvex} into
\begin{subequations}
\label{Problem:Relaxed} 
\begin{align}
 & \underset{\r,\T}{\text{minimize}}  
 & & \alpha_1 \|\r\|_1 + \alpha_2 \| \T \|_1 \\
 & \text{subject to} & & r_i \in [0,1],\, T_{ip} \in [0,1], \quad i\in\mathcal{V}_{\textrm{s}}, \, p \in\mathcal{V}  \\
& 
&& \eqref{eq.Tminw_cvx}, \eqref{eq.prob1}, \eqref{eq.flow} \\ 
& 
&& f(\r) \leq \gamma. 
\end{align}
\end{subequations}
With the assumption that the now continuous function $f: \Reals^J \to \Reals$ is convex in $\r$ (as it happens with all the aforementioned quality measurement examples~\cite{Joshi09}), then the program~\eqref{Problem:Relaxed} is convex. In addition, for the mentioned examples of $f(\r)$, \eqref{Problem:Relaxed} is a semidefinite program, which makes its solution efficient to compute polynomially with off-the-shelf software. We indicate with $(\hat{\r}, \hat{\T})$ any possible solution of~\eqref{Problem:Relaxed}. 

It is important to note that the couple $(\hat{\r}, \hat{\T})$ is only an approximation of the sought solution $(\r^*, \T^*)$. However, we will see in the simulation section that  $(\hat{\r}, \hat{\T})$ is usually a sparsely enough approximate solution. An additional feature of the approximate couple $(\hat{\r}, \hat{\T})$ is that it is feasible w.r.t. the constraint set of the original problem~\eqref{Problem.nonconvex}, and therefore it does not have to be mapped into a different set (as it usually happens in relaxed sensor selection problems). The reason for this is that we are working with rates and not Boolean variables. 

A strategy to increase the sparsity of the approximate couple $(\hat{\r}, \hat{\T})$, which has been proposed in~\cite{Candes08}, is to use a reweighted $\ell_1$ minimization mechanism. In this paper, we also use this strategy, which goes as follows. Consider the relaxed problem~\eqref{Problem:Relaxed}, with the different cost function $\alpha_1 \|\w\odot\r\|_1 + \alpha_2 \|\W\odot \T \|_1 $, where $\w \in \Reals^{J}$ and $\W \in \Reals^{J\times (J+K)}$ are a weighting vector and matrix, respectively. The weights can be determined so to push small components of $\r$ and $\T$ to zero, and boost big ones to one. In particular, initialize $w^0_i = 1$ and $W^0_{ip} = 1$, then for each $\tau\geq 0$ solve the problem
\begin{subequations}
\label{Problem:RelaxedWeighted} 
\begin{align}
 & \underset{\r,\T}{\text{minimize}}  
 & & \alpha_1 \|\w^{\tau}\odot\r\|_1 + \alpha_2 \|\W^{\tau}\odot \T \|_1 \\
 & \text{subject to} & & r_i \in [0,1],\, T_{ip} \in [0,1], \quad i\in\mathcal{V}_{\textrm{s}}, \, p \in\mathcal{V}  \\
& 
&& \eqref{eq.Tminw_cvx}, \eqref{eq.prob1}, \eqref{eq.flow} \\ 
& 
&& f(\r) \leq \gamma, \label{eq.MSErate}
\end{align}
\end{subequations}
whose solution is $(\hat{\r}^\tau, \hat{T}^\tau)$, and whose weights for $\tau \geq 1$ are $w^\tau_i = w^{\tau-1}_i/(\epsilon + \hat{r}^{\tau-1}_i)$ and $W^{\tau}_{ip} = W^{\tau-1}_{ip}/(\epsilon + \hat{T}^{\tau-1}_{ip})$, with $\epsilon$ a small positive constant. 

This iterative (reweighted) procedure delivers sparser solutions, as we will show in the simulation results. We have summarized the resulting sparse sensor and link selection (SSLS) iterative algorithm in Algorithm \ref{alg:SeLiS}.

{{\bf Connectivity guarantees of Algorithm~\ref{alg:SeLiS}.}
We notice that due to~\eqref{eq.flow}, the solution coming from Algorithm \ref{alg:SeLiS} guarantees that the measurements acquired at the sensors are delivered at the APs, i.e., each of the active sensors has a path back to at least one AP. To formally prove this statement, consider~\eqref{eq.flow}:
\begin{equation*}
r_i \bar{r}_i + \sum_{p\in\mathcal{V}_{\mathrm{s}}} T_{pi}R_{pi} \leq  \sum_{p\in\mathcal{V}}T_{ip}R_{ip}, \quad i \in \mathcal{V}_{\textrm{s}},
\end{equation*}
this constraint has to be true for each active sensor (the one for which $r_i>0$), and it reads $0 \leq 0$ for the not active ones (due to constraint~\eqref{eq.Tminw_cvx}, i.e., in this case also $T_{pi}$ and $T_{ip}$ are $0$). Since it has to be true for all active sensors, each of them has to send out more rate than what it receives (and the difference is given by its measurement rate), that is 
$$
\sum_{p\in\mathcal{V}_{\mathrm{s}}} T_{pi}R_{pi} <  \sum_{p\in\mathcal{V}}T_{ip}R_{ip}, \quad i \in \{j \in \mathcal{V}_{\textrm{s}}|r_j>0\}.
$$
Therefore, first: no active sensor can be a sink (it has to send out more than it receives). Second: there cannot be loops of active sensors not connected to a sink. In fact, if there were, since the rate augments along the loop, constraint~\eqref{eq.flow} would not be satisfied for at least a pair of active sensors connected together. Thus, the only possibility is that eventually each sensor has a path to a sink. This is also what we observe in simulations. \hfill $\Box$
}

\begin{algorithm}[t]
\footnotesize
\begin{algorithmic}[1] 
\REQUIRE Number of iterations $N$, reweighting tolerance $\epsilon>0$, sensor importance $\alpha_1\geq 0$, link importance $\alpha_2\geq 0$.
\STATE Set the weighting vector and matrix as $w^0_i = 1$ and $W^0_{ip} = 1$ for all $i\in \mathcal{V}_\textrm{s}$ and $p \in \mathcal{V}$
\FOR {$\tau = 0$ to $N-1$}
	\STATE Solve the convex program~\eqref{Problem:RelaxedWeighted} with off-the-shelf interior point methods (e.g., SDPT3~\cite{Toh1999} or SeDuMi~\cite{Sturm1998}). Let the solution be $(\hat{\r}^\tau, \hat{\T}^\tau)$. 
	\STATE Compute the new weights $\w^{\tau+1}$ and $\W^{\tau+1}$ as 
	$$
	w^{\tau+1}_i = \frac{w^{\tau}_i}{\epsilon + \hat{r}^{\tau}_i}, \quad W^{\tau+1}_{ip} = \frac{W^{\tau}_{ip}}{\epsilon + \hat{T}^{\tau}_{ip}}
	$$
\ENDFOR
\STATE Output the solution couple $(\hat{\r}^N, \hat{\T}^N)$
\end{algorithmic}
\caption{Sparse Sensor and Link Selection}
\label{alg:SeLiS}
\end{algorithm}

\begin{remark}\label{rem.distributed}
\emph{(Distributed algorithms)} Although the algorithms in this paper are centralized, one could devise distributed algorithms in a standard fashion. For instance, Problem (10) and (19) with the choice for $f(\r)$ of (2) fit the general structure presented in \cite{Simonetto2015}. In particular one needs to consider the local decision variables $x_i$ as the vector $(r_i, \{T_{ip}\}_{p\in \mathcal{V}_{\mathrm{s}}})$. In this case, with the use of consensus-based dual decomposition each sensor could decide their on/off strategy and to whom to communicate. Nonetheless, first, the re-weighting procedure is not trivial to implement in this case, and second, the sensors could spend a considerable amount of battery power to decide their on/off strategy. We believe that developing distributed and yet efficient (i.e., power-aware) algorithms for sensor selection is still an open research area, which is left for future investigations. 
\end{remark}

\begin{remark}\label{rem.stochastic}
{\emph{(Stochasticity of the reliability matrix $R_{ip}$)} Although here we assume to know each element $R_{ip}$ in a deterministic sense, one could also think of estimating $R_{ip}$ online. If then one possesses a pdf for $R_{ip}$, one could replace the deterministic constraint~\eqref{eq.flow} with a stochastic variant of it. Another approach in the estimation would be the one of~\cite{Kim2011a}. Finally, a third approach would consider a time-varying online algorithm to track $R_{ij}$ as it (possibly) varies in time, which is in line with the research proposed in~\cite{Paper1}.  
}
\end{remark}


\begin{remark}\label{rem.energy}
{\emph{(Energy efficiency)}   Energy efficiency can also be considered in the proposed approach. For instance, one could re-run the selection algorithm to take into account that the battery charge of the devices has changed, so to keep a balance in the usage of the whole sensor network. A way to include battery charge into the optimization problem is, e.g., to initialize the weights $w_i^0$'s not to $1$ but to the inverse of the battery level: $1$ if fully charged, $\infty$ if out of charge. 
}
\end{remark}


\section{Numerical Results for sensors and links}
\label{SimulationResults1}


In this section, we assess the performance of the proposed SSLS algorithm in terms of the amount of resources that are used, i.e., the number of both, active sensors and links. We also verify the consistency and the subgraph connectivity.

We consider an estimation scenario where sensors are randomly deployed according to a uniform distribution in a square area of side $5$ units. The regression matrix, $\A=\left[\textbf{a}_1, \cdots, \textbf{a}_J\right]^{\intercal}$ $\A \in \mathbb{R}^{J\text{x}m}$, is drawn from a zero-mean Gaussian distribution with variance $1$. 
The variance of the noise is the same at all sensors, $\sigma_i=1/ \sqrt{\text{SNR}}$, where SNR is set to $0$ dB. We use the cost $f(\r)$ of \eqref{CostFunction} and set the parameter $\gamma$ in \eqref{eq.MSErate} to 0.5. 
The link reliability metric that we use in the simulations is given by:

\begin{equation}
   \label{LinkReliability}
   R_{ip}= \left\{
     \begin{array}{ll}
      1 - \frac{1}{2}({\frac{\parallel \x_i - \x_p \parallel}{d}})^{2\beta}    & \mbox{if} ~ 0 \le  \parallel \x_i - \x_p \parallel < d    \\
      \frac{1}{2}(2-\frac{\parallel \x_i - \x_p \parallel}{d})^{2\beta}      &     \mbox{if} ~ d \le \parallel \x_i - \x_p \parallel < 2d \\
      0                                      &   \mbox{otherwise}  \\
   \end{array} \right.
\end{equation}
with $\beta$ the power attenuation factor ($2 \leqslant \beta\leqslant 6$) and $d$ the communication radius. We have considered $\beta=2$ and $d=1.74$ \cite{Arroyo-Valles07}.

The number of iterations in the reweighted $\ell_1$ minimization is empirically set to 30 to trade-off sparsity of the solution and computational time. 
Due to the application of the reweighted $\ell_1$ minimization mechanism, only the sensors and links with relatively high acquisition rate and link probability are active. We round off to 0 the link probabilities and sensor rates lower than a sufficiently small constant $\delta$, which is set to $\delta = 2\cdot 10^{-4}$. Further, we consider $\alpha_1=\alpha_2=1$. {Notice that rounding off the probabilities to $0$ could incur in a loss of connectivity. This is however not likely in practice, due to the reweighting procedure that makes sure that the non-zero probabilities have values well above the selected threshold $\delta$. The experimental results support this claim, since we have not witnessed any loss in connectivity.} 

\begin{figure}[tb]
	\centering
	\psfrag{a}{\small \text{y-axis}}
\psfrag{b}{\small \text{x-axis}}
			\includegraphics[scale=1]{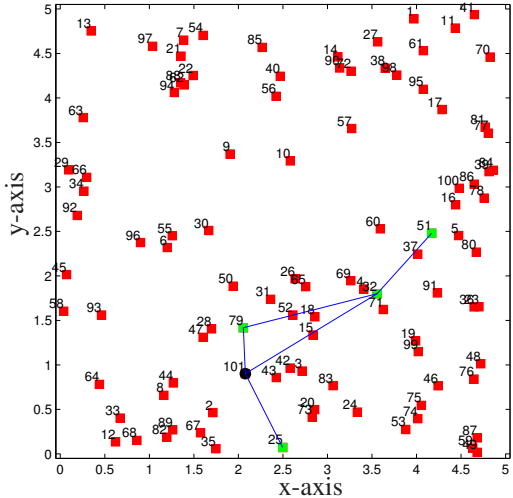} 
\caption[Network topology with selected sensors and links]{Active links and sensors in a one-AP network for $J=100$ nodes.}
	\label{fig:SLSelection}
\end{figure}

\begin{figure}[tb]
	\centering
		\psfrag{a}{\small \hskip1cm $\hat{r}_i$}
\psfrag{b}{\small \text{Sensor index $i$}}
			\includegraphics[scale=1]{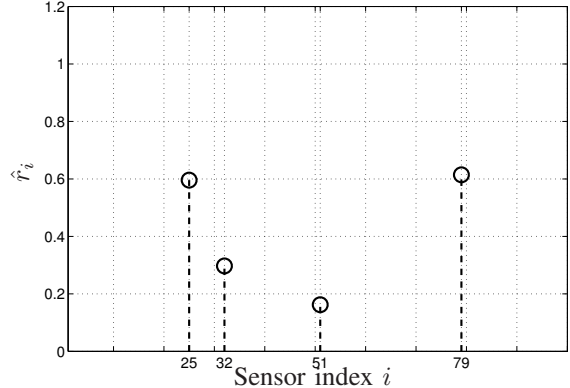}  
			\caption{Relative rates of the active sensors.}
	\label{fig:SLSelection_Index}
\end{figure}

 
Fig. \ref{fig:SLSelection} is an example of a 100-node sensor network with a single AP. The parameter to estimate has dimension $m=2$ and the maximum rate is $\bar{r}=0.7$. Active sensors are colored in green while the AP is in black. The results show the sparsity of the solution since only a few sensors (4\%) and links (0.072\%) are active. It can be also seen that the selected subgraph is connected and there is always a path between the active sensors and the AP. The solution also satisfies the other constraints of the optimization problem. Fig. \ref{fig:SLSelection_Index} shows the relative rates of the active sensors. 

Next, our purpose is to show average performance results. Hence, we run $250$ Monte Carlo simulations for each network configuration. The number of deployed sensors, $J$, varies from 30 to 100 and there is one AP. Two values of $\bar{r}$ are considered, 0.4 and 0.7. Simulations are run considering that $\bar{r}$ is identical for all nodes in the sensor network. Also, we consider two values for the dimension of the parameter to estimate, $m=2$ and $m=4$. 
Two metrics are considered to assess the performance of the networks. They try to measure the amount of resources that are used in the network. 

Since we are dealing with acquisition rates, let us first define the total relative acquisition rate of the whole network as the sum of the acquisition rates of the sensors in the network, i.e., $\sum_{i\in\mathcal{V}_{\textrm{s}}} \hat{r}_i^N$. In order to make the performance measurement independent of the number of sensors in the network, we define the percentage of the total relative acquisition rate of the whole network, $P_{\textrm{trr}}$, as 

 \begin{equation}
P_{\textrm{trr}}=\frac{\sum_{i\in\mathcal{V}_{\textrm{s}}} \hat{r}_i^N}{J} \cdot 100.
 \end{equation}
Recall that the relative acquisition rate $\hat{r}_i^N \in [0,1]$. Note that only the active sensors contribute to the sum since their acquisition rate is different from 0, so this measure gives us information about the percentage of active sensors.

Considering that $\sum_{p \in \mathcal{V}} \hat{T}_{ip}^N \leq 1$ for $i\in\mathcal{V}_{\textrm{s}}$, we next define the percentage of the aggregate network link probability, $P_{\textrm{alp}}$, as 

 \begin{equation}
P_{\textrm{alp}}=\frac{\sum_{i\in\mathcal{V}_{\textrm{s}}} \sum_{p \in \mathcal{V}} \hat{T}_{ip}^N}{J} \cdot 100.
 \end{equation}

Note that only active sensors and APs contribute to the sum, since the remaining link probabilities are 0. In this case, the metric is related to the percentage of active links in the network.
In both cases, the lower the metrics are, the fewer resources (in terms of active sensors and links) are used.


\begin{figure}[t]
\centering
\psfrag{a}{\small \hskip1cm $P_{\textrm{trr}}, P_{\textrm{alp}}$}
\psfrag{b}{\small \text{Number of Sensor Nodes}}
\psfrag{lg1}{\small \text{sensors}  $\bar{r}=0.4$}
\psfrag{lg2}{\small \text{links} $\bar{r}=0.4$}
\psfrag{lg3}{\small \text{sensors}  $\bar{r}=0.7$}
\psfrag{lg4}{\small \text{links} $\bar{r}=0.7$}
\includegraphics[scale=1]{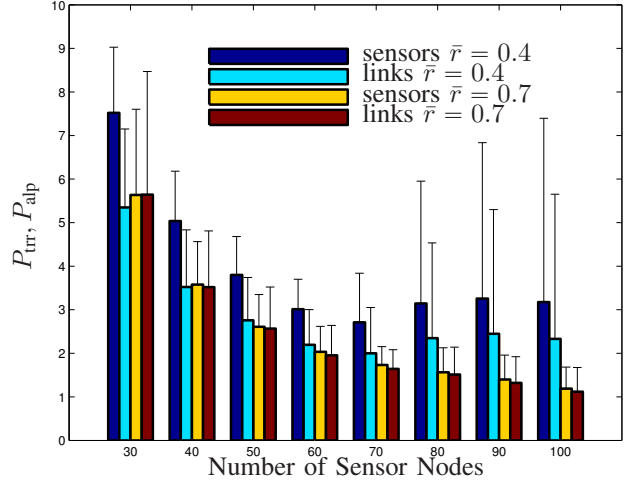} 
\caption {Average performance and its standard deviation for $m=2$ and for different amount of sensors and $\bar{r}$.}
	\label{fig:SLSelectionAvg_n2}
\end{figure}


\begin{figure}[t]
\centering
\psfrag{a}{\small \hskip1cm $P_{\textrm{trr}}, P_{\textrm{alp}}$}
\psfrag{b}{\small \text{Number of Sensor Nodes}}
\psfrag{lg1}{\small \text{sensors}  $\bar{r}=0.4$}
\psfrag{lg2}{\small \text{links} $\bar{r}=0.4$}
\psfrag{lg3}{\small \text{sensors}  $\bar{r}=0.7$}
\psfrag{lg4}{\small \text{links} $\bar{r}=0.7$}
\includegraphics[scale=1]{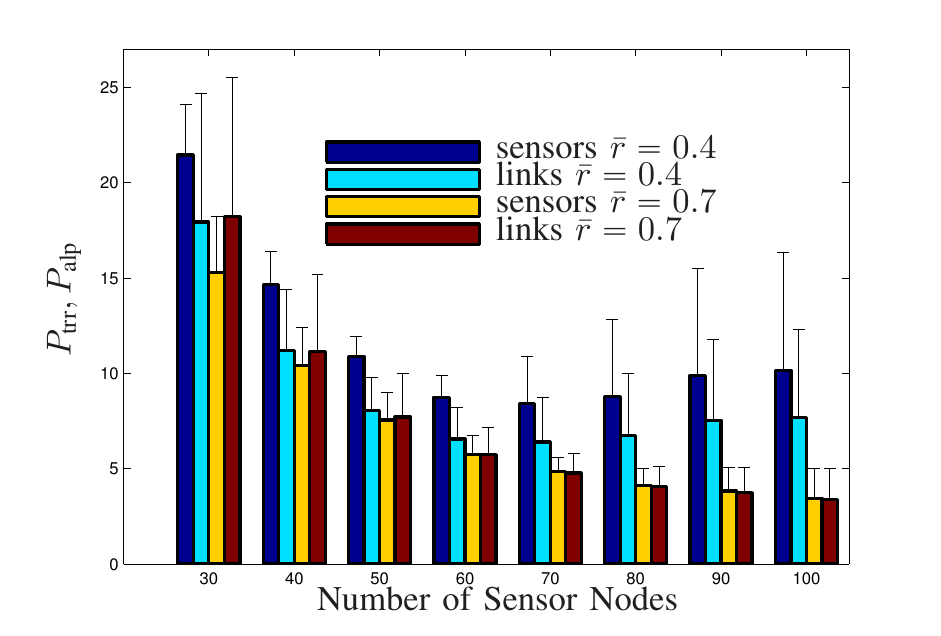} 
\caption{Average performance and its standard deviation for $m=4$ and for different amount of sensors and $\bar{r}$.}
	\label{fig:SLSelectionAvg_n4}
\end{figure}

\begin{figure}[t]
\centering
\psfrag{a}{\small \hskip-.8cm Percentage of active sensors / links}
\psfrag{b}{\small \text{Number of Sensor Nodes}}
\psfrag{lg1}{\small \text{sensors}  $\bar{r}=0.4$}
\psfrag{lg2}{\small \text{links} $\bar{r}=0.4$}
\psfrag{lg3}{\small \text{sensors}  $\bar{r}=0.7$}
\psfrag{lg4}{\small \text{links} $\bar{r}=0.7$}
\includegraphics[scale=1]{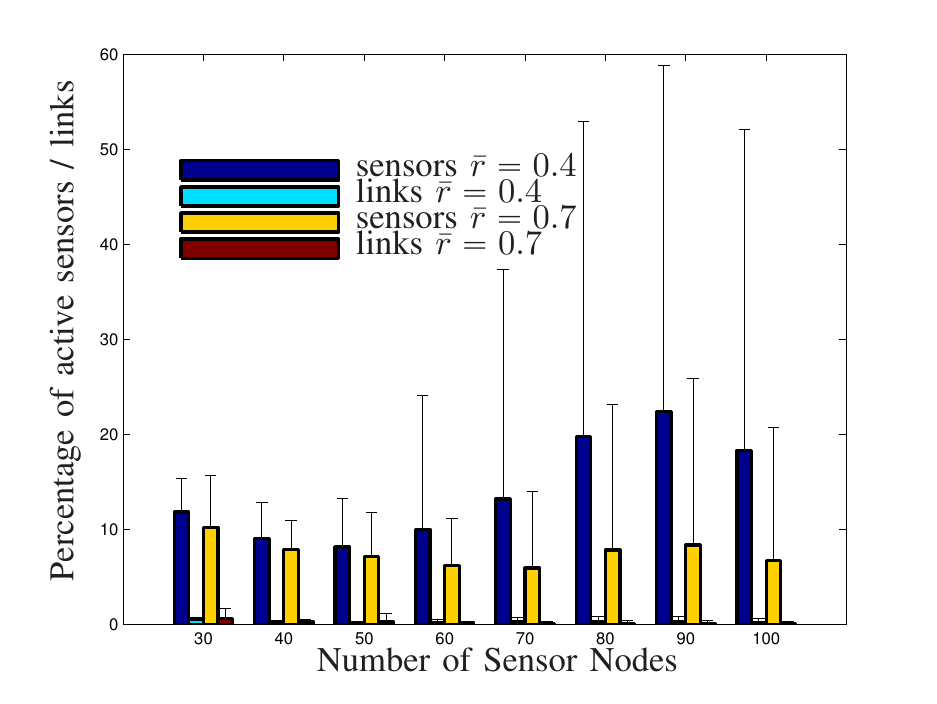}
\caption {Average percentage of active sensors and links and its standard deviation for $m=2$ and for different amount of sensors and $\bar{r}$.}
	\label{fig:SLSelectionAvg_n2_SLPerc}
\end{figure}

\begin{figure}[h!]
\centering
\psfrag{a}{\small \hskip-.8cm Percentage of active sensors / links}
\psfrag{b}{\small \text{Number of Sensor Nodes}}
\psfrag{lg1}{\small \text{sensors}  $\bar{r}=0.4$}
\psfrag{lg2}{\small \text{links} $\bar{r}=0.4$}
\psfrag{lg3}{\small \text{sensors}  $\bar{r}=0.7$}
\psfrag{lg4}{\small \text{links} $\bar{r}=0.7$}
\includegraphics[scale=1]{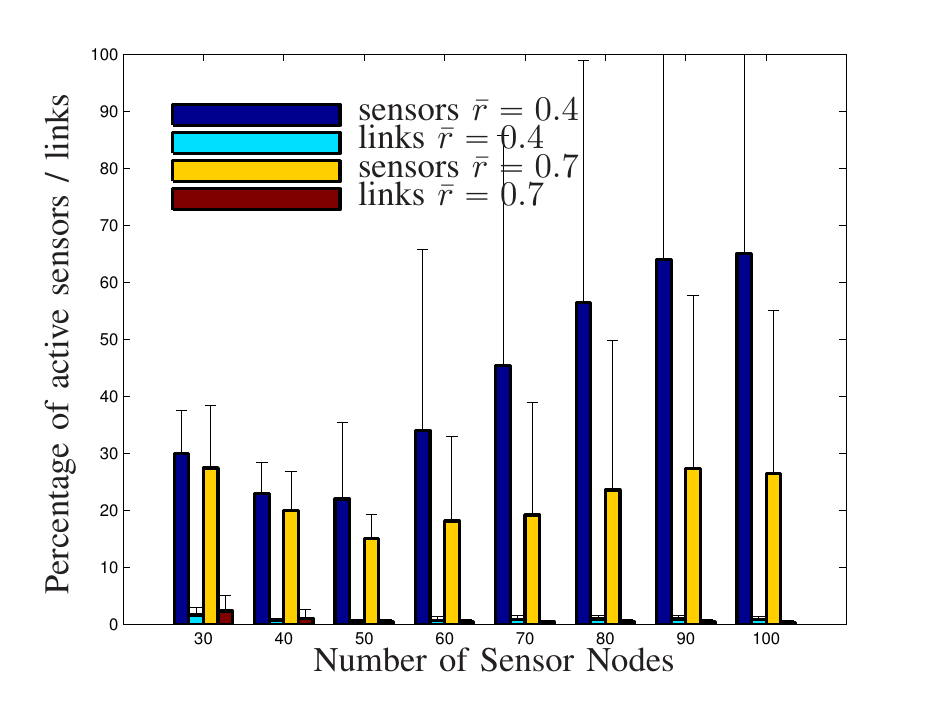}
 \caption {Average percentage of active sensors and links and its standard deviation for $m=4$ and for different amount of sensors and $\bar{r}$.}
	\label{fig:SLSelectionAvg_n4_SLPerc}
\end{figure}

Fig. \ref{fig:SLSelectionAvg_n2} and Fig. \ref{fig:SLSelectionAvg_n4} show the average performance and the standard deviation for $m=2$ and $m=4$, respectively, for different amounts of deployed sensors and the two values of $\bar{r}$. Even for the worst case scenario, i.e., for $\bar{r}=0.4$ and a $30$-node network, the $P_{\textrm{trr}}$ and $P_{\textrm{alp}}$ values are $8$\% and $5.5$\% for $m=2$ and $22$\% and $17$\% for $m=4$, respectively (which represents a small percentage of used resources). 

In order to verify if those metric values correspond to the activation of a low number of sensors with high relative rate values or correspond to a high number of active sensors with low relative rate values, Fig. \ref{fig:SLSelectionAvg_n2_SLPerc} and Fig. \ref{fig:SLSelectionAvg_n4_SLPerc} illustrate the average percentage of active sensors and links. For networks of $30$ nodes and $\bar{r}=0.4$, the average percentage of active sensors and links is $12$\% (i.e., $3.6$ sensors) and $0.55$\% for $m=2$ and $30$\% (i.e., $9$ sensors) and $1.7$\% for $m=4$, respectively. Thus, this result corroborates that the amount of used resources is conservative, i.e., there is a low percentage of active sensors with high relative rates because the $P_{\textrm{trr}}$ values are the highest in Fig. \ref{fig:SLSelectionAvg_n2} and Fig. \ref{fig:SLSelectionAvg_n4}. 

\subsection{Case $\bar{r}=0.7$ (yellow and red bars)}
From Fig. \ref{fig:SLSelectionAvg_n2} and Fig. \ref{fig:SLSelectionAvg_n4} it can be seen that the values of the two metrics decrease with the increase of the number of sensors (regardless of $m$), reaching lower values than those of the network of $30$ nodes. 

Let us examine the scenario with $m=2$ (analogous conclusions hold for networks with $m=4$). If we also  analyze the trend in the percentage of active sensors (Fig. \ref{fig:SLSelectionAvg_n2_SLPerc}), it first decreases and later increases slightly starting from networks of $80$ nodes. 
Even though networks with $80$ to $100$ nodes have between $6 \%$ to $8\%$ of active nodes (i.e., $7$ sensors), those networks have a slightly higher amount of active resources than in networks of $30$ nodes ($10\%$, approximately $3$ nodes). However, in general, the total number of active sensors stay low in comparison to the total number of sensors, which corroborates the sparsity of the solution.

\subsection{Case $\bar{r}=0.4$ (dark and light blue bars)}

Let us analyze the behavior of the metrics for $\bar{r}=0.4$ and $m=2$ (analogous conclusions are raised for networks with $m=4$). In Fig. \ref{fig:SLSelectionAvg_n2}, $P_{\textrm{trr}}$ values decrease from $7.5$\% at networks of $30$ nodes to $2.7$\% at networks of $70$ nodes, and from that point increases slightly up to a value of $3.2$\% at networks of $100$ nodes. If we now have a look at Fig. \ref{fig:SLSelectionAvg_n2_SLPerc}, the percentage of active sensors goes from $11.8$\% at $30$-node networks (i.e., $3.5$ sensors) to $8$\% at $50$-node networks (i.e., $4$ sensors) and later increases until reaching a value of $22$\% at $90$-node networks (i.e., $20$ nodes).
While the number of active nodes is similar in networks with a low amount of sensors ($30$ to $50$ nodes), it increases slightly for denser networks ($60$ to $100$ nodes). In this latter case, sensors are closer to each other so that the reliability values among sensors are similar and more sensors may be activated. First, although not reported here, we have observed that increasing the number of reweighting iterations does help in the latter case in reducing the amount of active sensors, at the cost of increasing the computational time requirements. Second, we will see how this is not an issue when relays are considered.

In case of the percentage of active links, the values are below $0.7$\% for $m=2$ and $2$\% for $m=4$ for all the network sizes. Hence, the networks are sparse in the amount of active links.

\vskip5mm

From the Figures, it can be appreciated that, for a given number of nodes, the percentage of used resources is lower in case of estimating a parameter of 2 dimensions than one of 4 dimensions (compare the metric values as well as the percentage of active sensors and links). Furthermore, the percentage of used resources (active sensors and links) is lower in case of considering $\bar{r}=0.7$.

Subgraph connectivity and consistency have been also checked for every run. All the activated sensors have a path to the AP. Furthermore, connectivity of the network in the sense of \eqref{eq.flow} is guaranteed.


\section{Sensor and Relay Selection}
\label{Relays}

When dealing with wireless sensor networks which are deployed in large areas, it is often useful to employ relays to facilitate the transmission of measurements back to the APs. In this spirit, we also consider the possible presence of relays. In particular, from here on, all the nodes deployed in the sensor network may act as sensors or as relays. Note that sensors can also act as a relay while sensing, as discussed in the previous section. Our goal is to consistently determine which of the nodes, which are placed at well-defined positions, should play the role of sensors and which ones the role of relays while guaranteeing both a prescribed network performance and connectivity in the selected  subgraph. Notice that relays have less energy requirements that sensors, and therefore the distinction between sensors and relays is beneficial to further reduce the overall energy consumption. Notice also that, as expressed in the introduction, the proposed solution may be reiterated in time, to assign different roles at different times. 


In order to model the possibility for a node to be acting as a sensor or as a relay, we introduce a new Boolean variable $\bnu \in \{0,1\}^{J+K}$, and we define that a node $i \in \mathcal{V}_{\textrm{s}}$, a sensor or relay, is on if $\nu_i = 1$ and it is off, otherwise ($\nu_p =1$ for APs). From the nodes that are on, we will know they are sensors when their $r_i$ is strictly positive, while the others are acting as relays. 

We also reformulate the constraints accordingly. The constraint~\eqref{eq.Tminw}  gets reformulated as 
\begin{equation}\label{eq.Tminnu}
T_{ip} \leq \min\{\nu_i, \nu_p\}, \quad i\in\mathcal{V}_{\textrm{s}}, p \in \mathcal{V},
\end{equation}
as the relays can exchange information. Notice that the constraint has a simplified form w.r.t.~\eqref{eq.Tminw}, since the variable $\bnu$ is Boolean. In addition, we need a constraint that makes sure that a sensor has a positive relative rate only when its node is activated, that is  
\begin{equation}\label{eq.wminnu}
r_i \leq \nu_i, \quad i\in\mathcal{V}_{\textrm{s}}.
\end{equation}
Finally, the constraint~\eqref{eq.flow} can be carried over as it is,
\begin{equation}\label{eq.flowrelay}
r_i \bar{r}_i + \sum_{p\in\mathcal{V}_{\mathrm{s}}} T_{pi}R_{pi} \leq  \sum_{p\in\mathcal{V}}T_{ip}R_{ip}, \quad i \in \mathcal{V}_{\textrm{s}},
\end{equation}

With this in place, the problem we want to solve is how to consistently select minimum rates, relays, and links so to guarantee a certain network performance and connectivity. We can formulate this as  
\begin{subequations}
\label{SensorRelayOptizProblemDeter0}  
\begin{align}
 & \underset{\r,\T,\bnu}{\text{minimize}}  
 & & \alpha_1 \|\r\|_0 + \alpha_2 \|\T\|_0 + \alpha_3 \|\bnu\|_0 \\
 & \text{subject to}& & r_i \in [0,1],\, T_{ip} \in [0,1], \nonumber \\ &&& \nu_i \in \{0,1\}, \quad i\in\mathcal{V}_{\textrm{s}}, \, p \in\mathcal{V}  \\
& 
&& \eqref{eq.prob1}, \eqref{eq.Tminnu}, \eqref{eq.wminnu}, \eqref{eq.flowrelay},  \\ 
& 
&& f(\r) \leq \gamma.   
\end{align}
\end{subequations}
Any possible solution of this problem is indicated as the triplet $(\r^*, \T^*, \bnu^*)$. This problem is a nonconvex mixed-integer programming problem and therefore finding any triplet $(\r^*, \T^*, \bnu^*)$ would be in general too computationally expensive. As done for the case where the relays are not present, we relax the problem to a convex one. In particular, we substitute the $\ell_0$ pseudo-norm with the convex surrogate $\ell_1$ norm, and we let the boolean vector $\bnu$ become real and live in the set $[0,1]^{J+K}$. With this, we arrive to the convex problem
\begin{subequations}
\label{SensorRelayOptizProblemDeter}  
\begin{align}
 & \underset{\r,\T,\bnu}{\text{minimize}}  
 & & \alpha_1 \|\r\|_1 + \alpha_2 \|\T\|_1 + \alpha_3 \|\bnu\|_1 \\
 & \text{subject to}& & r_i \in [0,1],\, T_{ip} \in [0,1], \nonumber \\ &&&  \nu_i \in [0,1], \quad i\in\mathcal{V}_{\textrm{s}}, \, p \in\mathcal{V}  \\
& 
&& \eqref{eq.prob1}, \eqref{eq.Tminnu}, \eqref{eq.wminnu}, \eqref{eq.flowrelay},  \\ 
& 
&& f(\r) \leq \gamma,   
\end{align}
\end{subequations}
whose solution is indicated with $(\hat{\r}, \hat{\T}, \hat{\bnu})$. Once again, the approximate triplet $(\hat{\r}, \hat{\T}, \hat{\bnu})$ is going to be different in general from the sought one $(\r^*, \T^*, \bnu^*)$. An important difference with problem~\eqref{Problem.nonconvex} and its relaxed version is the presence of the Boolean vector $\bnu$: this makes the triplet $(\hat{\r}, \hat{\T}, \hat{\bnu})$ in general unfeasible w.r.t. the constraints of the nonconvex problem~\eqref{SensorRelayOptizProblemDeter0} (the reason is that $\hat{\nu}_i$ does not have to be either $0$ or $1$). In this paper, we consider to project  $\hat{\nu}_i$ to $1$ any time $\hat{\nu}_i > 0$. In this way, the new triplet becomes feasible w.r.t. constraints of the nonconvex problem~\eqref{SensorRelayOptizProblemDeter0}.

In Algorithm~\ref{alg:SeReLiS}, we summarize the procedure for consistent sparse sensor, relay, and link selection (SSRLS), where we have used once again the sparse-enhancement procedure of reweighting. 

\begin{algorithm}[t]
\footnotesize
\begin{algorithmic}[1] 
\REQUIRE Number of iterations $N$, reweighting tolerance $\epsilon>0$, sensor importance $\alpha_1\geq 0$, link importance $\alpha_2\geq 0$, relay importance $\alpha_3\geq 0$.
\STATE Set the weighting vectors and matrix as $w^0_i = 1$, $v^0_i = 1$, and $W^0_{ip} = 1$ for all $i\in \mathcal{V}_\textrm{s}$ and $p \in \mathcal{V}$
\FOR {$\tau = 0$ to $N-1$}
	\STATE Solve the convex program
	\begin{align*}
 & \underset{\r,\T,\bnu}{\text{minimize}}  
 & & \alpha_1 \|\w^{\tau}\odot\r\|_1 + \alpha_2 \|\W^{\tau}\odot\T\|_1 + \alpha_3 \|\v^{\tau}\odot\bnu\|_1 \\
 & \text{subject to}& & r_i \in [0,1],\, T_{ip} \in [0,1], \nonumber \\ &&& \nu_i \in [0,1], \quad i\in\mathcal{V}_{\textrm{s}}, \, p \in\mathcal{V}  \\
& 
&& \eqref{eq.prob1}, \eqref{eq.Tminnu}, \eqref{eq.wminnu}, \eqref{eq.flowrelay},  \\ 
& 
&& f(\r) \leq \gamma,   
\end{align*}
	  with off-the-shelf interior point methods (e.g., SDPT3~\cite{Toh1999} or SeDuMi~\cite{Sturm1998}). Let the solution be $(\hat{\r}^\tau, \hat{\T}^\tau, \hat{\bnu}^\tau)$. 
	\STATE Compute the new weights $\w^{\tau+1}$, $\v^{\tau+1}$ and $\W^{\tau+1}$ as 
	$$
	w^{\tau+1}_i = \frac{w^{\tau}_i}{\epsilon + \hat{r}^{\tau}_i}, \quad W^{\tau+1}_{ip} = \frac{W^{\tau}_{ip}}{\epsilon + \hat{T}^{\tau}_{ip}}, \quad v^{\tau+1}_i = \frac{v^{\tau}_i}{\epsilon + \hat{\nu}^{\tau}_i}
	$$
\ENDFOR
\STATE Project $\hat{\nu}_i^N$ to $1$, if $\hat{\nu}_i^N> 0$.
\STATE Output the solution triplet $(\hat{\r}^N, \hat{\T}^N, \hat{\bnu}^N)$
\end{algorithmic}
\caption{Sparse Sensor, Relay, and Link Selection}
\label{alg:SeReLiS}
\end{algorithm}

{{\bf Connectivity guarantees of Algorithm~\ref{alg:SeReLiS}.} 
We formalize now the claim that from each active sensor there exists a path (formed by relays and other active sensors) that goes to an AP. The argument that we use to prove this claim is the same as the one that we have used to prove the connectivity guarantees of Algorithm~\ref{alg:SeLiS} (where no relay were considered). Consider~\eqref{eq.flowrelay}: this constraint has to be true for each active sensor and active relay (the one for which $r_i = 0$ and $\nu_i>0$), and it reads $0 \leq 0$ for the not active ones (due to constraints~\eqref{eq.Tminnu}-\eqref{eq.wminnu}). Since it has to be true for all active sensors and relays, the sensors have to send out more rate than what they receive (and the difference is given by $r_i \bar{r}_i$), while the relays can send out exactly what they receive. Therefore, first: no active sensor or relay can be a sink. Second: there cannot be loops containing active sensors not connected to a sink, since the rate augments along the loop and~\eqref{eq.flowrelay} would not be satisfied for at least one pair of connected active elements (either sensor-sensor, sensor-relay, or relay-relay). Third: there cannot be loops containing only active relays. The reason for the last claim is that, although~\eqref{eq.flowrelay} would be satisfied along the loop for any $T_{ip} = T_{pi}$ for all pairs of active relays $(i,p)$ on the loop, the solution $T_{ip}=0$ is the optimal one, given the selected cost function. Which induces all the relays in the loop to become inactive.  Thus, the only possibility is that eventually each sensor has a path to a sink, and no relays are used without purpose. \hfill $\Box$
}


\section{Numerical Results with Relays}
\label{SimulationResults2}

Similarly to the sensor and link selection scenario, in this section we assess the performance of the new SSRLS algorithm in terms of the amount of resources that are used. We also verify the consistency and the subgraph connectivity.

\begin{figure}[tb]
	\centering
		\psfrag{a}{\small \text{y-axis}}
\psfrag{b}{\small \text{x-axis}}
		\includegraphics[scale=1]{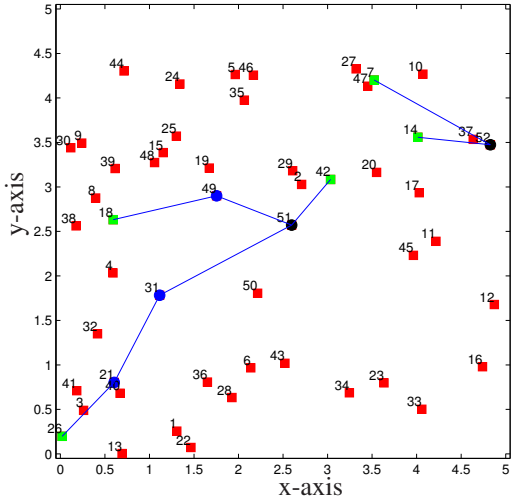}
	\caption{Selected sensors, relays and links in a two-AP 50-sensor network where $m=4$. }
	
	\label{fig:SRLSelectionSc1}
\end{figure}

Once again, we consider an estimation scenario, where the parameters are the same\footnote{In particular, the number of iterations in the reweighted $\ell_1$ minimization is set to 30 while $\delta = 2\cdot 10^{-4}$.}  as those used in the sensor and link selection case (Section \ref{SimulationResults1}). We consider $\alpha_1=\alpha_2=\alpha_3=1$.
Fig. \ref{fig:SRLSelectionSc1} is an example of a 50-node sensor network with two APs with $m=4$ and $\bar{r}=0.4$. The active sensors are colored in green, the active relays in blue and the APs are colored in black. Looking at the figure, it is evident that the obtained solution is sparse. From the 50 nodes (excluding the APs), 5 are selected as sensors and 3 as relays. The amount of active links (i.e., those that have a probability value higher than 0) is $8$\%. Observe the connectivity of the selected subgraph, where there is a path from each active sensor to the APs via the relays, where messages are routed stochastically according to the link probability. The solution also satisfies the other constraints of the optimization problem.

Next, and following a parallel analysis to the one made in the sensor and link selection scenario, we show average performance results. The number of sensors varies from 30 to 100, $\bar{r}$ is 0.4 or 0.7 and $m$ is either 2 or 4. 250 Monte Carlo simulations are run for each network configuration. The metrics to assess the network performance are the ones exposed in Section \ref{SimulationResults1}. To check the sparsity in the number of relays, we also evaluate the percentage of active relays in the network.

\begin{figure}
\centering
\psfrag{a}{\small $P_{\textrm{trr}}, P_{\textrm{alp}}$}
\psfrag{b}{\small \text{Number of Sensor Nodes}}
\psfrag{lg1}{\small \text{sensors}  $\bar{r}=0.4$}
\psfrag{lg2}{\small \text{links} $\bar{r}=0.4$}
\psfrag{lg3}{\small \text{sensors}  $\bar{r}=0.7$}
\psfrag{lg4}{\small \text{links} $\bar{r}=0.7$}
\includegraphics[scale=1]{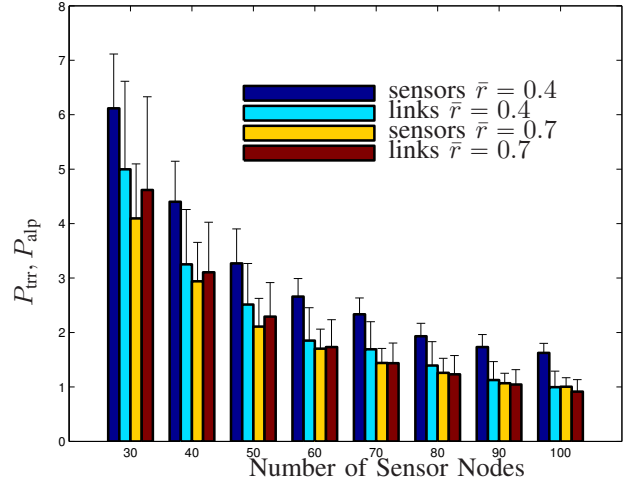}
 \caption {Average performance and its standard deviation for $m=2$ and for different amount of sensors and $\bar{r}$.}
	\label{fig:SLRelSelectionAvg_n2}
\end{figure}

\begin{figure}
\centering
\psfrag{a}{\small $P_{\textrm{trr}}, P_{\textrm{alp}}$}
\psfrag{b}{\small \text{Number of Sensor Nodes}}
\psfrag{lg1}{\small \text{sensors}  $\bar{r}=0.4$}
\psfrag{lg2}{\small \text{links} $\bar{r}=0.4$}
\psfrag{lg3}{\small \text{sensors}  $\bar{r}=0.7$}
\psfrag{lg4}{\small \text{links} $\bar{r}=0.7$}
\includegraphics[scale=1]{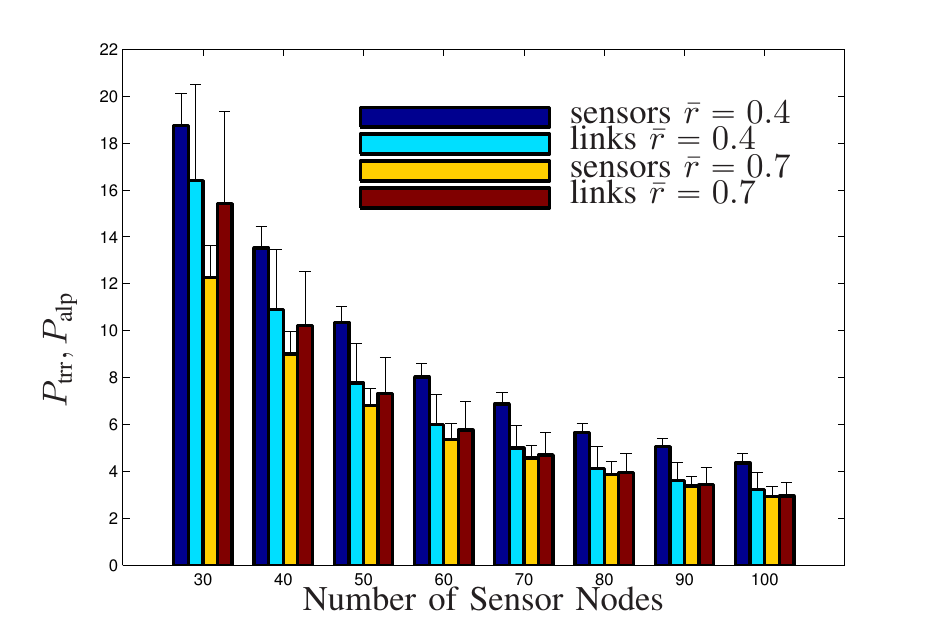}
 \caption {Average performance and its standard for $m=4$ and for different amount of sensors and $\bar{r}$.}
	\label{fig:SLRelSelectionAvg_n4}
\end{figure}

\begin{figure}
\centering
\psfrag{a}{\small \hskip-1cm Percentage of active sensors / links}
\psfrag{b}{\small \text{Number of Sensor Nodes}}
\psfrag{lg1}{\small \text{sensors}  $\bar{r}=0.4$}
\psfrag{lg2}{\small \text{links} $\bar{r}=0.4$}
\psfrag{lg3}{\small \text{sensors}  $\bar{r}=0.7$}
\psfrag{lg4}{\small \text{links} $\bar{r}=0.7$}
\includegraphics[scale=1]{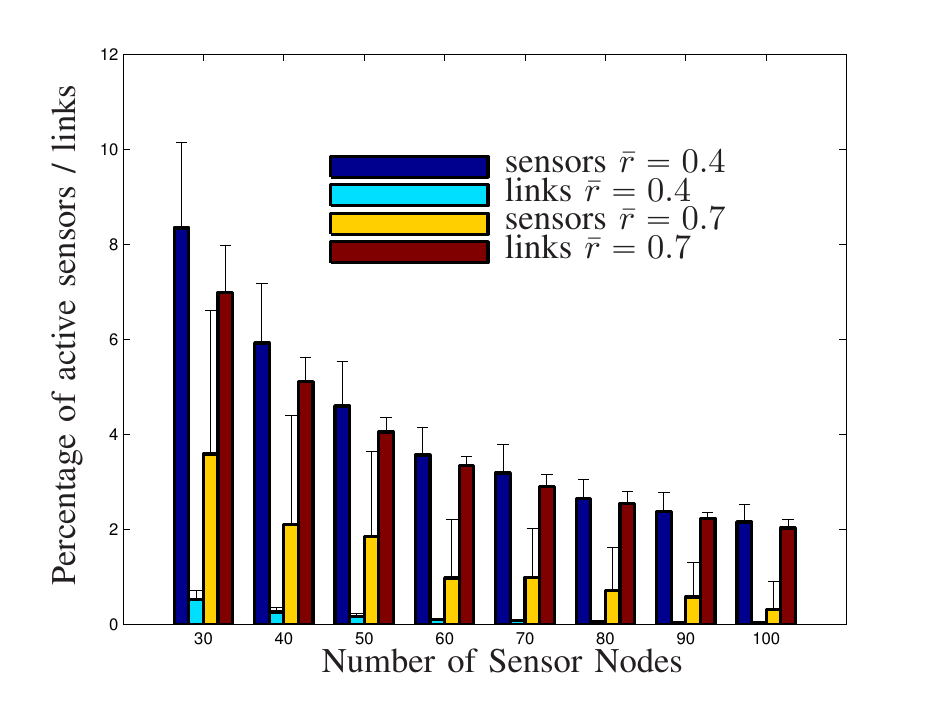}
\caption {Average percentage of active sensors and links and its standard deviation for $m=2$ and for different amount of sensors and $\bar{r}$.}
	\label{fig:SLRelSelectionAvg_n2_SLPerc}
\end{figure}

\begin{figure}
\centering
\psfrag{a}{\small  \hskip-1.5cm  Percentage of active sensors / links}
\psfrag{b}{\small \text{Number of Sensor Nodes}}
\psfrag{lg1}{\small \text{sensors}  $\bar{r}=0.4$}
\psfrag{lg2}{\small \text{links} $\bar{r}=0.4$}
\psfrag{lg3}{\small \text{sensors}  $\bar{r}=0.7$}
\psfrag{lg4}{\small \text{links} $\bar{r}=0.7$}
\includegraphics[scale=1]{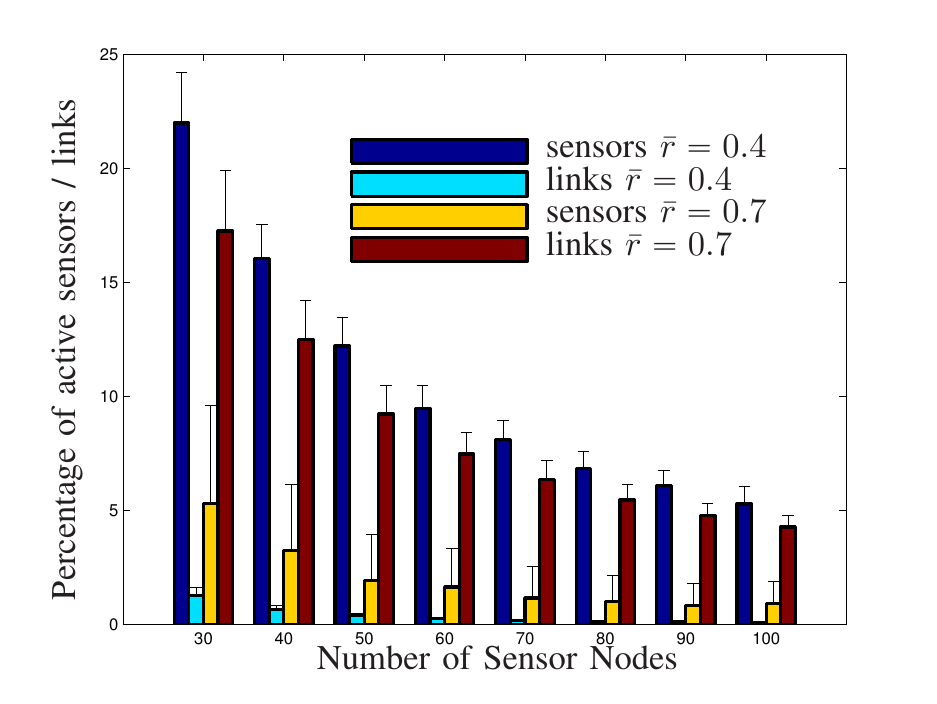} 
\caption {Average percentage of active sensors and links and its standard deviation for $m=4$ and for different amount of sensors and $\bar{r}$.}
	\label{fig:SLRelSelectionAvg_n4_SLPerc}
\end{figure}

\begin{figure}
\centering
\psfrag{a}{\small \hskip-.5cm \text{Percentage of active relays}}
\psfrag{b}{\small \text{Number of Sensor Nodes}}
\psfrag{lg1}{\small $\bar{r}=0.4, m=2$}
\psfrag{lg2}{\small $\bar{r}=0.7, m=2$}
\psfrag{lg3}{\small $\bar{r}=0.4, m=4$}
\psfrag{lg4}{\small $\bar{r}=0.7, m=4$}
\includegraphics[scale=1]{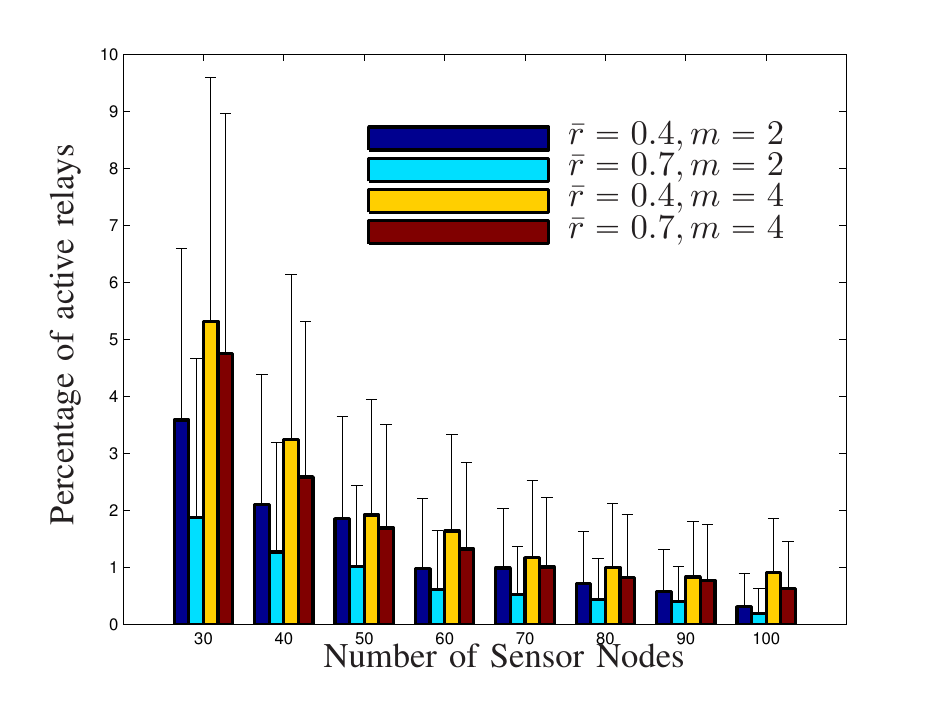} 
\caption {Average percentage of active relays and its standard deviation for different amount of sensors, $\bar{r}$ and $m$.}
	\label{fig:SLRelSelection_AvgRel}
\end{figure}

Fig. \ref{fig:SLRelSelectionAvg_n2} and Fig. \ref{fig:SLRelSelectionAvg_n4} show the average performance and the standard deviation, for $m=2$ and $m=4$, respectively, for different numbers of deployed sensor nodes and the two values of the maximum acquisition rate. 
{From these figures it can be seen that the $P_{\textrm{trr}}$ and $P_{\textrm{alp}}$ values decrease when the number of sensor nodes increases, going from a value of 6.3$\%$ ($m=2$, $\bar{r}=0.4$) or 19$\%$ ($m=4$, $\bar{r}=0.4$) for networks of $30$ nodes to values lower than 2$\%$ ($m=2$, $\bar{r}=0.4$) or 5$\%$ ($m=4$, $\bar{r}=0.4$) for networks of $100$ nodes. 
To verify if those metric values correspond to the activation of a few sensors with high relative rate value, Fig. \ref{fig:SLRelSelectionAvg_n2_SLPerc} and Fig. \ref{fig:SLRelSelectionAvg_n4_SLPerc} illustrate the average percentage of active sensors and links. For $\bar{r}=0.4$, the average percentage of sensors goes from approximately $8\%$ (i.e., $2.4$ sensors for $m=2$) or $22\%$ (i.e., $6.5$ sensors for $m=4$) in 30-node networks to around $2\%$ (i.e., $2$ sensors for $m=2$) or $5\%$ (i.e., $5$ sensors for $m=4$) in 100-node networks, respectively. 
Fig. \ref{fig:SLRelSelection_AvgRel} also illustrates the percentage of active relays for sensor networks of different sizes, $\bar{r}$ and $m$. For $m=4$ and $\bar{r}=0.4$, $2$ relays  (or 5.5$\%$ of the nodes) are active in 30-sensor networks, while $1$ relay ($0.9\%$) is active in 100-sensor networks.

The conclusions from these figures are two-fold: 
First, independently of the total number of available nodes, the algorithm robustly selects a similar number of sensors, relays and links to satisfy the constraint on the measurement errors. This strongly suggests that the optimality of the sensing, given the constraints, is achieved. 
Second, the active sensors are those with high relative rates. And even more, we obtain sparse solutions not only in the amount of active sensors but also in the amount of active links and relays.

As observed for the sensor and link selection scenario, the demand of resources (percentage of active sensors, links and relays) is less when considering higher maximum rates (See Fig. \ref{fig:SLRelSelectionAvg_n2_SLPerc} , Fig. \ref{fig:SLRelSelectionAvg_n4_SLPerc} and Fig. \ref{fig:SLRelSelection_AvgRel}). Also, for a given number of nodes, the amount of used resources grows whenever the dimension of the parameter to estimate increases.
Besides, the variability in the results of the sensor, relay and link selection problem is lower than in the sensor and link selection problem. This can be observed by taking into account the standard deviation in the figures. All in all, it appears that when one considers also the presence of relays, one obtains better performance in terms of reduced active resources than in the case of no relays. A more in depth characterization is left as future research.}

Therefore, the SSRLS algorithm provides a consistent solution to the sensor and relay selection problem by always finding a connected path among the active sensors, relays and APs no matter the size of the network and the dimension $m$ of the parameter to estimate, and which satisfies the network performance constraint for the active sensors. 

However, the following question may arise: are the active sensors obtained after solving the SSRLS algorithm the same as those that would be active in a problem which aims at selecting the minimum number of sensors and their corresponding acquisition rates to satisfy a certain MSE-rate, i.e., solve the relaxed version of the following problem: $$\underset{\r}{\text{minimize}} \quad  \|\r\|_0 \quad \text{subject to} \quad f(\r) \leq \gamma,$$ where sensors that satisfy $r_i>\delta$ (have an acquisition rate different from 0) are selected?

\begin{figure}
	\centering
{
		\psfrag{a}{\small \text{y-axis}}
\psfrag{b}{\small \text{x-axis}}
		\includegraphics[scale=1]{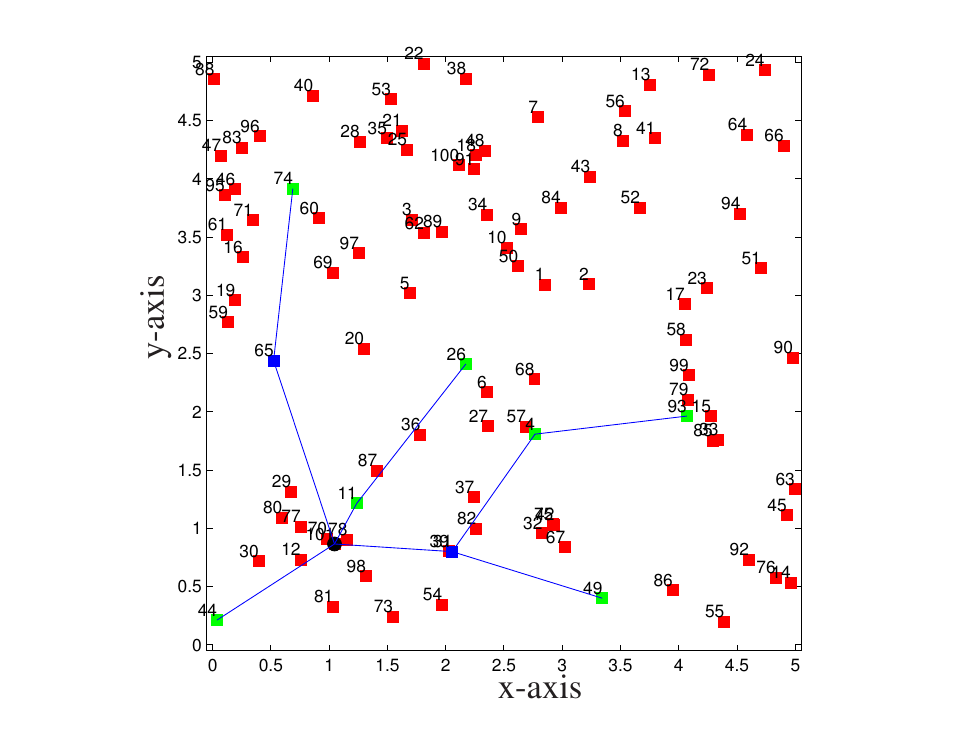} 
}
	\caption{Selected sensors, relays and links in a one-AP 100-sensor network where $m=4$ and $\bar{r}=0.7$. }
	\label{fig:SRLSel_comp1}
\end{figure}

\begin{figure}
	\centering
		\psfrag{a}{\small \text{y-axis}}
\psfrag{b}{\small \text{x-axis}}
		\includegraphics[scale=1]{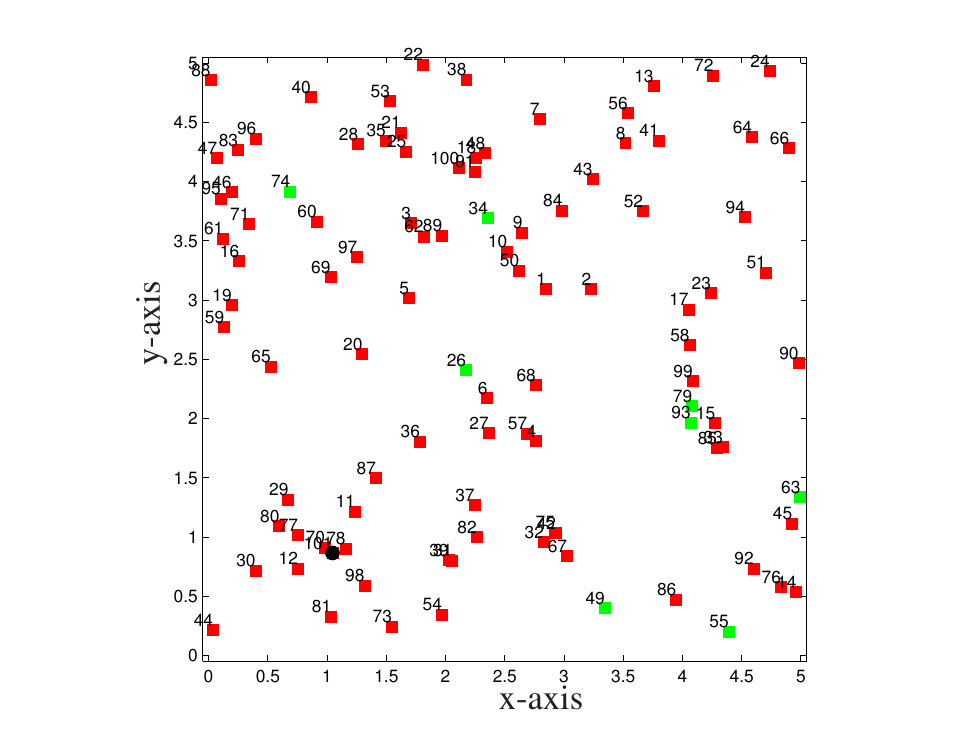} 
	\caption{Selected sensors in a one-AP 100-sensor network where $m=4$ and $\bar{r}=0.7$. }
	\label{fig:SLSel_comp1}
\end{figure}



Clearly, the answer is no. As an example, compare the active sensors in Fig.\ref{fig:SRLSel_comp1} and Fig. \ref{fig:SLSel_comp1}, for a one-AP 100-sensor network where $m=4$ and $\bar{r}=0.7$. The solution provided by the SSRLS algorithm not only takes into account the sensors with the highest acquisition rates (to satisfy the MSE-rate constraint), but also selects in a robust, coherent and consistent way relays and probability links such that the active sensors are connected to the APs (it considers the sensor deployment, too). On the contrary, the solution provided by the sensor selection problem does not consider the spatial distribution of sensors, and the only issue that matters is the selection of the sensors with the best acquisition rates. Obviously, this does not mean that both solutions do not activate some common sensors. In the previous example, sensors with indexes 26, 49, 74 and 93 are selected in both solutions.


\section{Link selection}
\label{SpecialCasesExtensions}


This last scenario is a particular case of the general one where we assume that all sensors are active, acquire measurements with relative rate at least $r_{i0}$, and communicate with the APs in a multi-hop fashion. The problem that remains is to determine the probabilistic routing matrix $\T$, that selects the minimum subset of links so that a certain constraint is satisfied. In particular, we want to ensure network integrity, defined according to \cite{Zavlanos13} as the ability of the network to support the desired communication rates in a certain network topology. 

As in the general scenario, the network needs to satisfy the flow inequality constraint given by \eqref{eq.flow}  in order to guarantee that messages are delivered to the APs. Furthermore, it is also required that sensors communicate their measurements with the APs at a nominal rate of $r_{i0}$ messages per time unit. This means that the relative acquisition rate should satisfy the following inequality: $r_i \geq  r_{i0}$. Thus, we aim at finding the appropriate relative rates $\r \in [0,1]^J$ and the sparse probabilistic routing matrix $\T$.


The network's objective function to be optimized is the social utility value of the optimization variables, $U_i(r_i)$ for the relative rate $r_i$,  and $V_{ip}(T_{ip})$ for the links $T_{ip}$, which is defined as $\sum_{i=1}^J U(r_i) + \sum_{i=1}^J \sum_{p=1}^{J+K} V_{ip}(T_{ip})$. 
Following \cite{Zavlanos13}, we measure the utility value associated to the rate as $U(r_i)= log(r_i) $, which penalizes small rates $r_i$, and the utility value of the links as $V_{ip}(T_{ip})=-T_{ip}^2$.
 
Then, the optimization problem that we have to solve is given by\footnote{Note that these utilities can be also incorporated in the objection functions of the previous optimization problems. However, and for the sake of simplicity, we only consider them in this scenario. }
\begin{subequations}
\label{LinkOptizProblemV0} 
\begin{align}
 & \underset{\r,\T}{\text{maximize}}  
 & &  \sum_{ i \in \mathcal{V}_{\textrm{s}}} U(r_i) + \sum_{ i \in \mathcal{V}_{\textrm{s}}} \sum_{p\in\mathcal{V}} V_{ip}(T_{ip})  - \alpha \| \T \|_0 \\
 & \text{subject to} & & r_i \in [0,1],\, T_{ip} \in [0,1], \quad i\in\mathcal{V}_{\textrm{s}}, \, p \in\mathcal{V}  \\
& 
&& \eqref{eq.prob1}, \eqref{eq.flow} \\ 
& 
&& r_i \geq  r_{i0},  \quad i \in \mathcal{V}_{\textrm{s}}, 
\end{align}
\end{subequations}

The problem in \eqref{LinkOptizProblemV0} is not convex due to the $\ell_0$-norm in the objective function. Thus, we relax the non-convex term of \eqref{LinkOptizProblemV0} by substituting the 
$\ell_0$-pseudo norm, with the $\ell_1$ norm. Then, the previous optimization problem is transformed into the following one

\begin{subequations}
\label{LinkOptizProblem} 
\begin{align}
 & \underset{\r,\T}{\text{maximize}}  
 & &  \sum_{ i \in \mathcal{V}_{\textrm{s}}} U(r_i) + \sum_{ i \in \mathcal{V}_{\textrm{s}}} \sum_{p\in\mathcal{V}} V_{ip}(T_{ip})  - \alpha \| \T \|_1 \\
 & \text{subject to} & & r_i \in [0,1],\, T_{ip} \in [0,1], \quad i\in\mathcal{V}_{\textrm{s}}, \, p \in\mathcal{V}  \\
& 
&& \eqref{eq.prob1}, \eqref{eq.flow} \\ 
& 
&& r_i \geq  r_{i0},  \quad i \in \mathcal{V}_{\textrm{s}}, 
\end{align}
\end{subequations}

The objective function is strictly concave and the constraints are linear inequalities, so the problem can be solved efficiently by using convex optimization tools. Note that the optimal utility depends on the spatial configuration of the sensors, and consequently the optimal link probabilities and rate variables do too, which are denoted as $r_{\textbf{x},i}^*$, and $T_{\textbf{x},ip}^*$.

The amount of selected links depends on parameter $\alpha$, which controls the sparsity level (the higher it is, the fewer links are selected). In order to increase sparsity and avoid the tuning of parameter $\alpha$, we apply the iterative reweighted  $\ell_1$ minimization algorithm only to the third term of the objective function (we call this partial reweighted $\ell_1$ minimization), which diminishes the influence of that parameter and helps in the link selection process. We round off to 0 the link probabilities lower than a sufficiently small constant $\delta$.

\begin{figure}
\centering
\psfrag{a}{\small \text{y-axis}}
\psfrag{b}{\small \text{x-axis}}
\psfrag{lg1}{\footnotesize \text{sensors}}
\psfrag{lg2}{\footnotesize \text{sink}}
\psfrag{lg3}{\footnotesize \text{$\delta \leq T_{i,j} < 0.25$}}
\psfrag{lg4}{\footnotesize \text{$0.25 \leq T_{i,j} < 0.5$}}
\psfrag{lg5}{\footnotesize \text{$0.5 \leq T_{i,j} < 0.75$}}
\psfrag{lg6}{\footnotesize \text{$0.75 \leq T_{i,j} \leq 1$}}
\includegraphics[scale=1, trim=0cm 0 0 0, clip=on]{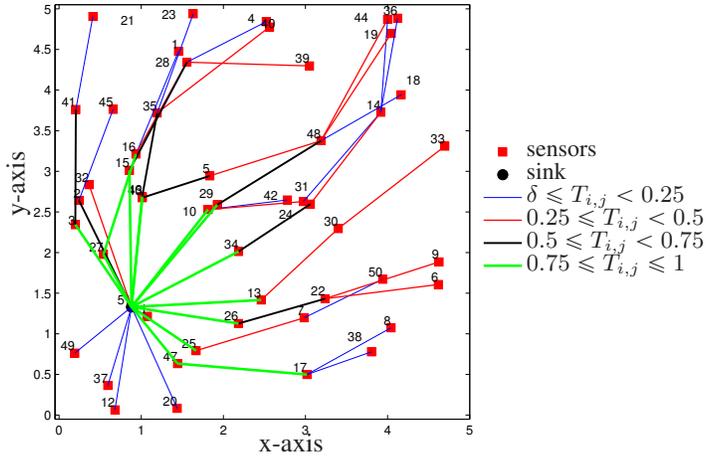} 
\caption {Selected links in a one-AP 50-node network or $\alpha=1$, and using partial reweighted $l_1$ . Sensors are colored in red, the AP in black and different colors in the links represent the different probability ranges.}
	\label{fig:LinkSelExam}
\end{figure}

In the remaining of this section, we show the performance of the link selection scheme. Fig. \ref{fig:LinkSelExam} is an example of a 50-node sensor network with one AP. Sensors are colored in red and the AP in black. The nominal rate is $r_{i0}=0.2$, which is identical for all the sensor nodes; another weighting parameter is $\alpha=1$. $\delta$ and the rest of the parameters are identical to those defined in Section \ref{SimulationResults1}. The color and the thickness of the links are related to the routing probability values, which are graded into different ranges. Blue links have a probability value between $\delta$ and $0.25$ and their line is the finest. The red ones have a probability between $0.25$ and $0.5$, the black links between $0.5$ and $0.75$ and the green ones between $0.75$ and $1$, having the thickest line. 

Note that only $51$ links are active (i.e., it is a sparse solution). Every sensor is connected via multiple hops (links that connect the sensors) to the AP, where the links with higher probabilities are always established between the AP and some of its neighboring nodes. This is logical given that a message that has been routed through multiple sensors should have a higher probability of arriving successfully to the AP. In general, the sensors which are placed far from the AP tend to establish links with low probability values. 

\begin{figure}
\centering
\psfrag{a}{\small \hskip0cm\text{Percentage of active links}}
\psfrag{b}{\small \text{Number of Sensor Nodes}}
\psfrag{lg1}{\footnotesize \text{Total - $\delta \leq T_{i,j} \leq 1$}}
\psfrag{lg2}{\footnotesize \text{$\delta \leq T_{i,j} < 0.25$}}
\psfrag{lg3}{\footnotesize \text{$0.25 \leq T_{i,j} < 0.5$}}
\psfrag{lg4}{\footnotesize \text{$0.5 \leq T_{i,j} < 0.75$}}
\psfrag{lg5}{\footnotesize \text{$0.75 \leq T_{i,j} \leq 1$}}
\includegraphics[scale=1]{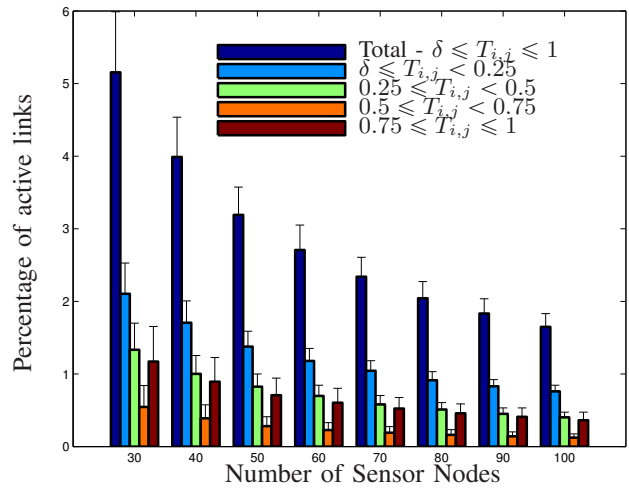} 
\caption {Average percentage of total active links and its standard deviation for different number of sensor nodes, and when link probabilities are graded in different ranges. }
	\label{fig:LSelectionAvg_n2}
\end{figure}

Fig. \ref{fig:LSelectionAvg_n2} illustrates the average percentage of active links (the total percentage and the percentage by probability ranges) for sensor networks composed of a number of sensors whose amount varies between 30 to 100 and $r_{i0}=0.2$. 250 Monte Carlo simulations are run for each network configuration. First, the figure shows the low amount of active links, so that the matrix $\T$ is sparse. For 30-node networks, the average percentage of active links is slightly higher than 5 $\%$. The percentage values decrease whenever the number of nodes in the network increases. Regarding the different probability ranges, the highest percentage of active links corresponds to values of $T_{ij}$ between $\delta$ and 0.25, and it is followed by links with probabilities $T_{ij}$ between 0.75 and 1. As in the earlier example, most of those links are established between the AP and its neighbors, which ensures that messages arrive to the AP.   




\section{Conclusions}

In this paper we have proposed two optimization methods for selecting optimally and consistently the minimum set of sensors (and their corresponding sensing rates) and links (and their link probability values); or sensors, relays and links, in wireless sensor networks. The chosen scenario has been parameter estimation, where the selected sensors have to guarantee a prescribed network performance based on the MSE-rate.  
Numerical results showed the sparsity of the solution, which translates into a smart use of the network resources. The proposed algorithms have provided a consistent solution to the selection problem by always finding a connected path among the selected set of sensors, relays and APs no matter the size of the network and the dimension of the parameter to estimate. This ensures the compliance of the network performance constraint by the selected sensors.  

Future work will consider the study of these algorithms from a decentralized point of view, eliminating the need of having an AP that collects all the measurements. The application of these algorithms to other scenarios is also a matter of further studies.

\footnotesize
\bibliographystyle{IEEEtran}
\bibliography{../../TeX/PaperCollection2,bib}

\end{document}